\RequirePackage{fix-cm}
\documentclass[smallextended]{svjour3}       
\smartqed  
\usepackage{graphicx}
\begin{document}

\title{The quest for $\mu 	\to e \gamma$: present and future}


\author{Francesco Renga}


\institute{F.~Renga \at
              Istituto Nazionale di Fisica Nucleare -- Sez. di Roma \\
              P.le A.~Moro 2, 00185 Roma\\
              \email{francesco.renga@roma1.infn.it} 
}

\date{Received: date / Accepted: date}

\maketitle

\begin{abstract}
The quest for $\mu \to e \gamma$ is one of the most important endeavors to search for New Physics beyond the
Standard Model. In this talk I will review the current status of the experimental searches by the MEG
Collaboration at PSI. I will also present a study of the experimental limiting factors that will define
the ultimate performances, and hence the sensitivity, in the search for $\mu \to e \gamma$ with continuous
muon beams of extremely high rate (one or even two orders of magnitude larger than the present
beams), whose construction is under consideration for the next decade.
\end{abstract}

\section{Introduction}
\label{intro}

Lepton flavor conservation is an accidental symmetry in the Standard Model (SM), not related to the gauge structure of the model but merely arising from 
its particle content, namely the absence of right-handed neutrinos. As a consequence, most of New Physics (NP) models predict some Lepton 
Flavor Violation (LFV) effects and, indeed, they are already strongly constrained by the present limits, like 
$BR(\mu \to e \gamma) < 4.2 \times 10^{-13}$ from the MEG experiment~\cite{meg}. The search for charged LFV is hence a clean and effective 
way to search for NP.

 
The $\mu^+ \to e^+ \gamma$ decay is searched for in the decay at rest of stopped muons from a high-intensity continuous beam. Both the
electron and the positron will have an energy of 52.8~MeV and will be emitted collinearly and back-to-back. A prompt background
comes from the radiative muon decay $\mu^+ \to e^+ \nu_e \overline \nu_\mu \gamma$, when the neutrinos carry a very small 
fraction of the available energy. Nonetheless, the dominant background at very high intensities is the accidental coincidence
of a positron from a normal muon decay and a photon from the radiative decay of another muon or the annihilation in flight
of another positron. The observables which can be used to discriminate signal and background are hence the positron and photon energies, 
the relative angle of their directions and their relative time.


A $\mu \to e \gamma$ experiment is composed of a thin target to stop muons, a positron section able to
reconstruct the positron momentum and trajectory and give a precise timing, and a photon section with very good
energy, time and position resolutions. 

According to~\cite{kuno-okada}, the accidental background rate depends on the beam intensity and resolution according to:
\begin{equation}
\Gamma_{\mathrm{acc}} \propto \Gamma_\mu^2 \cdot \delta E_e \cdot (\delta E_\gamma)^2 \cdot \delta T_{e\gamma} \cdot (\delta \Theta_{e\gamma})^2
\end{equation}
where $\Gamma_\mu$ is the muon stopping rate and $\delta E_e$, $\delta E_\gamma$, $\delta T_{e\gamma}$ and $\delta \Theta_{e\gamma}$ are
the energy, time and angular resolutions. The dependence on the square of $\Gamma_\mu$ makes useless a beam intensity increase if the total
background yield over the experiment lifetime is not negligible. Under these conditions, it can be advantageous to loose some efficiency if it allows
to improve the resolution, and recover the efficiency loss by increasing the beam rate, which is otherwise not possible.

The search for $\mu \to e \gamma$ relies on the availability of high-intensity muon beams like the ones delivered at the Paul Scherrer Institut (PSI) in 
Switzerland, with up to $10^8$ muons per second. It could be possible with the present technologies to increase this intensity by one or two orders of 
magnitudes. A study have been performed~\cite{future_meg} to identify the experimental factors which would limit the sensitivity of future searches 
for $\mu \to e \gamma$ with beams of such a high intensity.

\section{The status of the MEG-II experiment}

\begin{figure}
\centering
\includegraphics[width=0.6\textwidth]{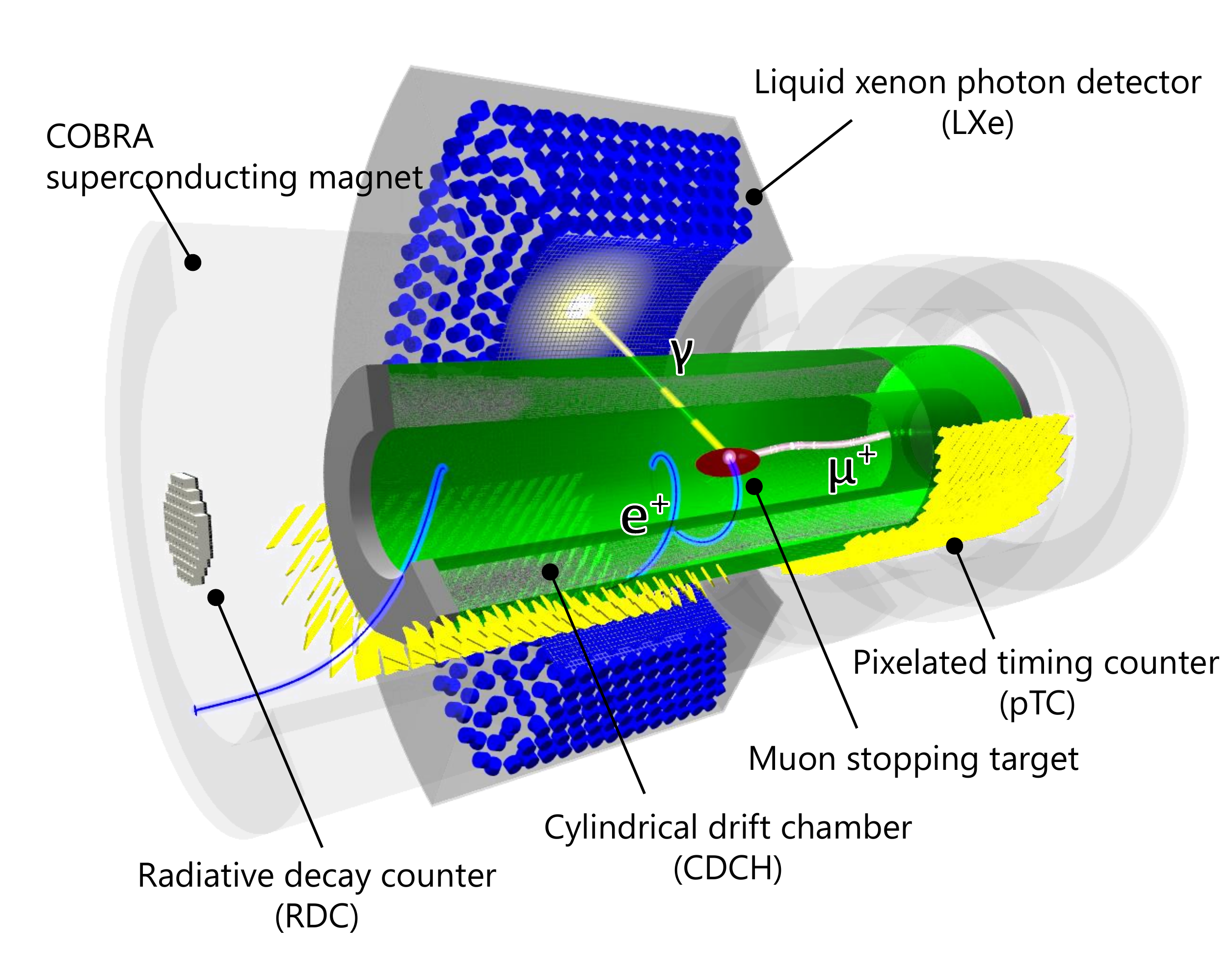}
\caption{The MEG II experiment }
\label{fig:meg2}       
\end{figure}

The MEG collaboration is currently finalizing an upgrade of all sub-detectors, with the goal of improving by one order of magnitude the
sensitivity reached in the first phase of the experiment. Figure~\ref{fig:meg2} shows a scheme of the new experiment, MEG II~\cite{megII}. 
Like MEG, it is composed of a positron spectrometer in a non-uniform magnetic field, a positron timing detector (\emph{Timing Counter}) and
a LXe calorimeter for the photon detection.

The construction of the new Timing Counter has been completed in 2017. The detector is made of 512 scintillator tiles
read out by SiPM. Positrons go through several tiles whose time measurements are combined. It requires an accurate calibration 
of the time offsets between the different tiles, performed by means of a dedicated laser system and using positron tracks from muon decays.
Both techniques have been tested in an engineering run in 2017 and the detector already reached the goal resolution of $\sim 35$~ps.

The 16 planar drift chamber that composed the MEG positron spectrometer will be replaced by a single cylindrical drift chamber with 
9 layers of stereo wires. 
The single hit resolution of the new chamber is expected to be $\sim 100~\mu$m. The resulting positron momentum resolution 
will be improved by a factor of 3 with respect to MEG, going below 100~keV/c, while the positron reconstruction efficiency will be improved 
by a factor of 2, mainly thanks to the longer extent of the chamber. The wiring and sealing of the drift chamber has been completed 
in Summer 2018 and the chamber will be tested on beam at PSI at the end of the year.

The LXe calorimeter of the MEG experiment has been upgraded by replacing the photomultiplier tubes in the in the inner face of the
detector with MPPCs customarily designed in collaboration with Hamamatsu in order to improve their sensitivity to the UV light emitted
by LXe. The MPPCs allow a better coverage of the inner surface of the calorimeter, with a significative improvement of the energy and
position resolution, in particular for photons converting just at the entrance of the calorimeter. An average energy resolution of 1\% 
at 52.8~MeV is expected. The first photons in the upgraded calorimeter have been detected in the 2017 engineering run.

A radiative muon decay veto, composed of LYSO crystals and plastic scintillators, will be added in MEG II to identify 
events with a low energy positron in coincidence with a high energy photon making background in the calorimeter.

The largely increased number of readout channels of the new experiment stimulated the development of a new data acquisition scheme, which
integrates trigger and data acquisition capabilities in a single system. Prototypes of the acquisition electronics have been successfully
tested in the 2016 and 2017 engineering runs.

The MEG detector is expected to be tested on beam, for the first time with all sub-detectors, in the second half of 2018, and to start taking physics
data in 2019, for a 3-year run. The improved resolutions will allow to increase the muon beam rate up to $7 \times 10^7$ muons per second,
compared to $3 \times 10^7$ in MEG. The MEG II experiment is foreseen to reach an expected upper limit of $6 \times 10^{-14}$ on 
the BR of $\mu \to e \gamma$.

\section{Future high-intensity muon beams}
\label{sec:beams}

Experiments searching for $\mu \to e \gamma$ with muon beams exploit the production of muons by a proton beam impinging on a target.
Protons produce pions that decay and give muons. 
The most intense continuous muon beams are delivered at PSI, with intensities up to $10^8$ muons per second. The
laboratory is considering the possibility of building a new beam line with an increased muon collection efficiency at the production target
and an increased transport efficiency toward the experimental areas. It should be possible to reach a rate of $10^{10}$ muons per second~\cite{himb}.

This rate is limited by the thickness of the production target. At PSI it stops 12-18\% of the protons in the beam, which needs
to be preserved to serve a neutron spallation source downstream of the muon production target. An alternative approach is being
explored at RCNP in Osaka, Japan, with the MuSIC project~\cite{music}. A thicker target is used in this case, allowing to increase by two orders of magnitude
the muon yield per unit of proton beam power. Although the projected muon beam rate is lower than what can be obtained at PSI, it is a good
demonstration of an alternative approach which can be used to reach unprecedented intensities.

The construction of a continuous muon beam line is also under consideration in the context of the PIP-II project~\cite{pipII} at Fermilab, USA. The
goal is to reach intensities similar to what could be obtained at PSI.

\section{Experimental techniques, limiting factors and sensitivity for $\mu \to e \gamma$ at future muon beams}

Due to the $\Gamma^2_\mu$ dependence of the accidental background rate, an increased muon beam intensity can be exploited in 
the search for $\mu \to e \gamma$ only if the detector resolutions can be improved so that the background yield is kept at a negligible level.
Hence, it is important to identify the factors which will limit the resolutions of the next generation of $\mu \to e \gamma$ experiments. 

A magnetic spectrometer complemented with fast detectors is almost an obliged choice for the positron detection. In the MEG-II drift chamber, 
the multiple scattering will already give a significative contribution to the resolutions. Hence, gaseous detectors are the preferred choice for 
the spectrometer, owing to their low material budget. The unavoidable material in the muon stopping target and in front of the detector will
finally limit the angular resolutions to about 4~mrad. Also, the momentum resolution expected in MEG-II is likely to be irreducible even with the
best compromises of resolutions and low material budget. When high intensity muon beams are considered, the detector aging also plays a crucial role.
One of the main technological issues for future experiments will be to face this issue, and replacing gaseous detectors with solid state detectors could
be unavoidable, with a consequent deterioration of the performances.

For the photon, two options can be considered: the calorimetric and the photon-conversion approach. A calorimeter provides very high efficiency, mostly limited by the interaction of the photon with the material in front of the calorimeter, with good energy, position and time resolutions. The 
photon-conversion approach exploits $e^+e^-$ pair creation in a thin conversion layer, followed by tracking of the pair in a spectrometer. It 
can give extremely good resolutions but a very poor efficiency. At very high beam intensity, anyway, this approach can still be competitive, as discussed above.

For calorimetry, the choice of the scintillating material determines the performances of the detector. A very good candidate could be LaBr$_3$(Ce), which
could allow to reach an energy resolution of 800~keV at 52.8~MeV with an extremely good time resolution (30~ps). This material is very expensive
but it could allow nonetheless to significantly increase the acceptance of the photon detector with respect to the one of MEG and MEG-II ($\sim 10\%$),
which has to cope with the extremely high cost of Xenon. 

For the conversion technique, the performances are determined by the pair production probability and the fluctuations of the 
energy loss of the $e^+e^-$ pair in the conversion material. The best compromise is obtained for high-$Z$ materials, like Lead and Tungsten. 
A resolution of 800~keV and an efficiency of  3\% is obtained for a converter thickness of 0.1 radiation lengths. It is important to notice that the 
photon conversion technique also allows to get a rough estimate of the photon direction, that helps to reject the accidental background. On the 
other hand, this technique needs to be complemented with fast detectors for timing. If one wants to stack multiple conversion layers, they need 
to be interleaved with fast detectors of lower $Z$ and it deteriorates the performances of the system. An alternative could be the implementation 
of an active conversion layer using fast and thin silicon detectors~\cite{TT-PET}.

There is also some room for an optimization of the target. In particular, the possibility of using multiple targets can be considered because, if the
conversion technique allows to determine the target where the photon has been produced, it allows to reduce the accidental background. 

\begin{figure}
\centering
\includegraphics[width=0.45\textwidth]{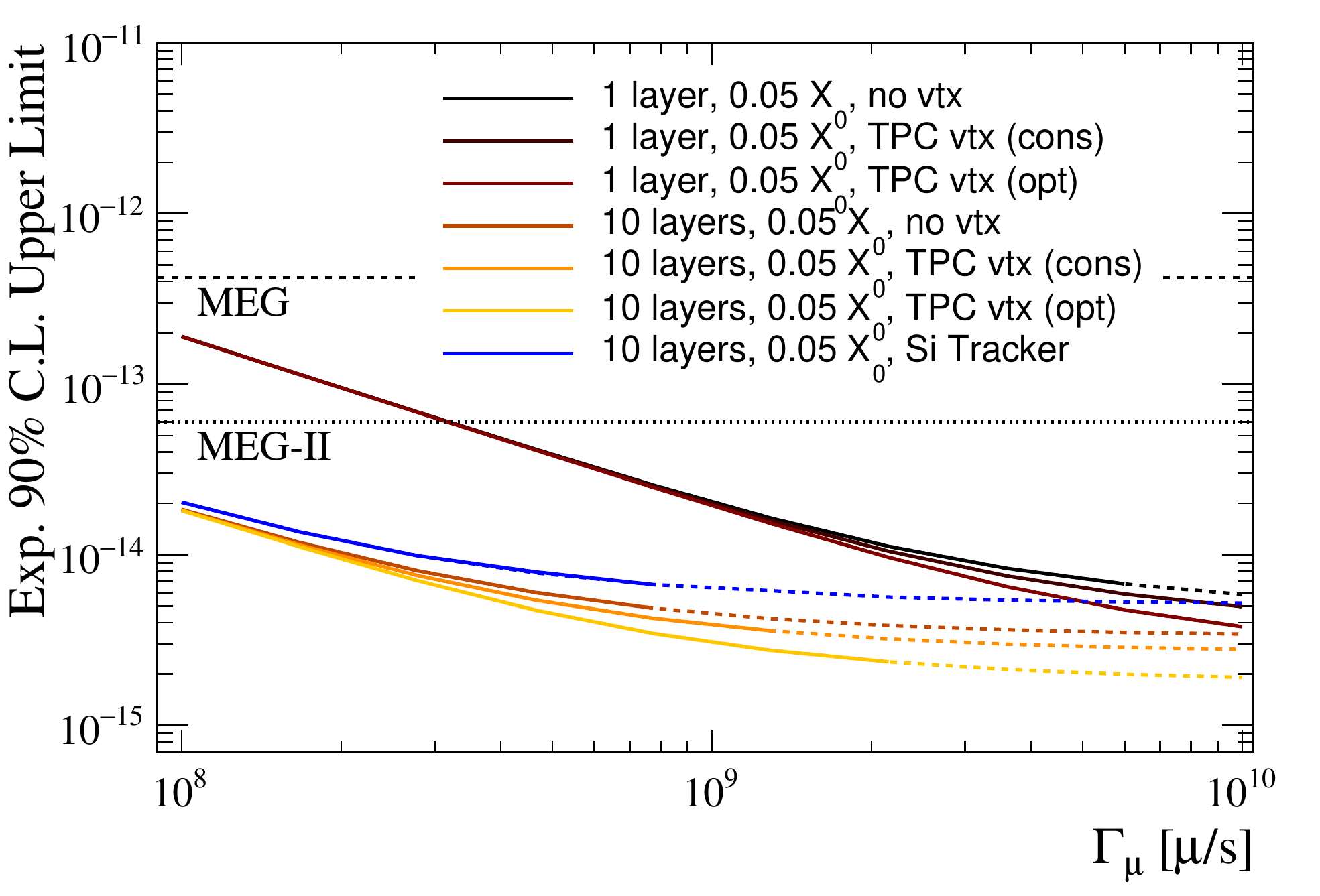}
\includegraphics[width=0.45\textwidth]{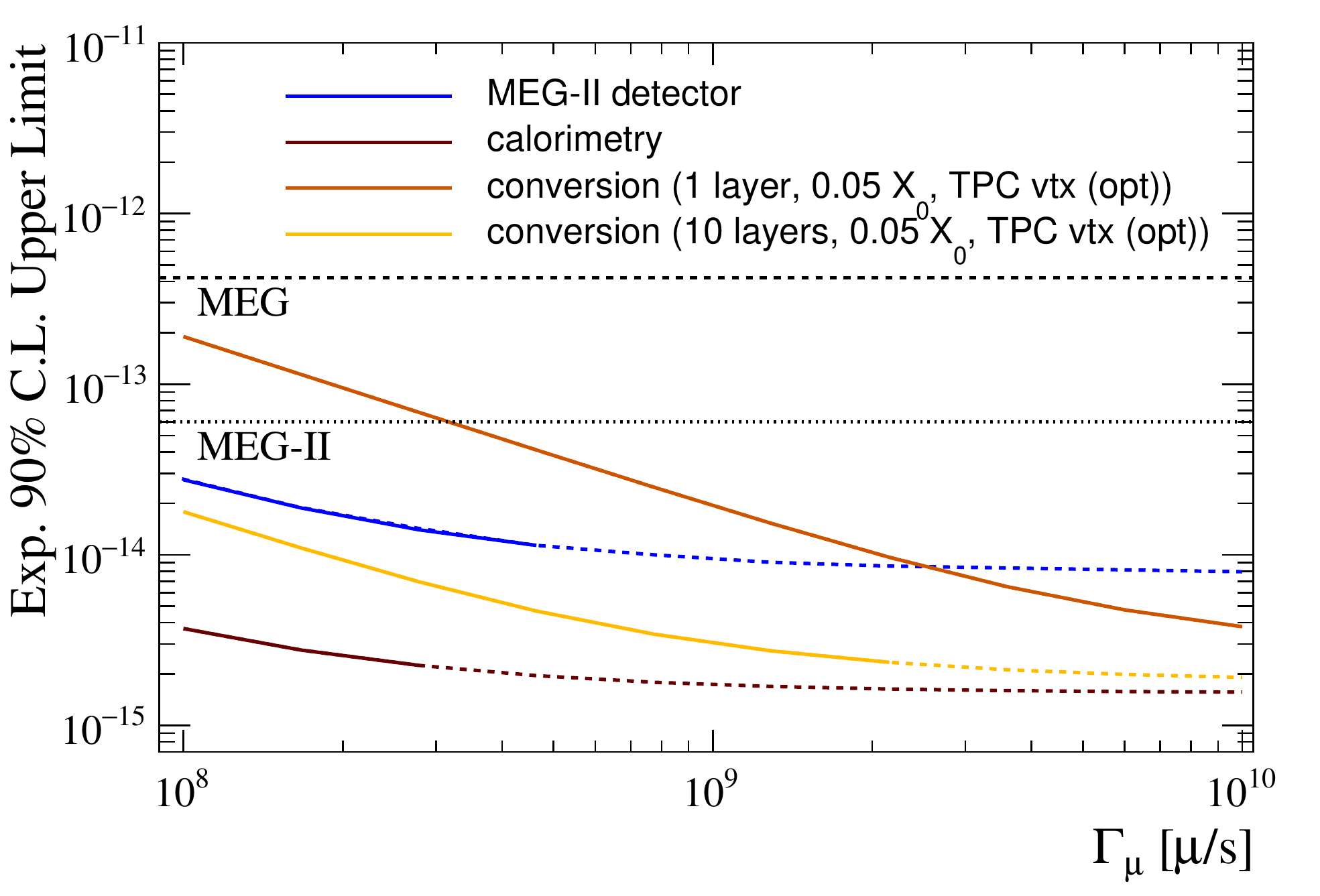}
\caption{Expected 90\% C.L. upper limit on the BR of $\mu \to e \gamma$ in different scenarios for a 3-year run. A few different 
designs are compared, including a TPC vertex detector option under conservative and optimistic hypotheses. 
The lines turn from continuous to dashed when the number of background events exceeds 10. The horizontal dashed and dotted 
lines show the current MEG limit and the expected MEG-II sensitivity, respectively.}
\label{fig:sens}       
\end{figure}

We imagined a conceptual experiment to search for $\mu \to e \gamma$ assuming reasonable incremental improvements in the detector technologies,
taking into account the limiting factors discussed above. Expected upper limits on the BR of $\mu \to e \gamma$
have been evaluated assuming a counting experiment and making use of the Feldman-Cousins algorithm~\cite{FC} under different scenarios.
The results are shown in Fig.~\ref{fig:sens} as a function of the muon beam intensity. 

\section{Conclusions}

The search for LFV is one of the most promising field in the quest for NP. The present limit on $\mu \to e \gamma$ by the MEG 
collaboration already strongly constrains the NP models and an improvement of one order of magnitude is expected with MEG-II. 
We investigated some long term prospects for the $\mu \to e \gamma$ search. Our estimates show that a 3-year run with an 
accelerator delivering 
around $10^9$ muons per second could allow to reach a sensitivity of a few $10^{-15}$ (expected 90\% upper limit 
on the $\mu \to e \gamma$~BR), with poor perspectives of going below $10^{-15}$ even with $10^{10}$ muons per second. 
Below $5 \times 10^8$ muons per second, the calorimetric approach needs to be used in order to reach this target. 
If a muon beam rate exceeding $10^9$ muons per second is available, the much cheaper photon conversion option 
would be recommended and would provide similar sensitivities.

The sensitivity would be eventually limited by the fluctuations of the interaction of the particles with the detector materials: 
this indicates that a further step forward in the search for $\mu \to e \gamma$~would require a radical  rethinking of the 
experimental concept.


\end{document}